Beyond Reductionism Twice: No Laws Entail Biosphere Evolution, Formal Cause Laws Beyond Efficient Cause Laws

Stuart Kauffman March 20, 2013

Introduction

Newton set the stage for our view of how science should be done. We remain in what I will call the "Newtonian Paradigm in all of physics, including Newton, Einstein, and Schrodinger. As I will show shortly, Newton invented and bequeathed to us "efficient cause entailing laws" for the entire becoming of the universe. With Laplace this became the foundation of contemporary reductionism in which all that can happen in the world is due to efficient cause entailing laws. More this framework stands as our dominant way to do science. The Newtonian Paradigm has done enormous work in science, and helped lead to the Industrial Revolution, and even our entry into Modernity.

In this paper I propose to challenge the adequacy of the Newtonian Paradigm on two ground: 1) For the evolution of the biosphere beyond the watershed of life, we can formulate no efficient cause entailing laws that allow us to deduce the evolution of the biosphere. A fortiori, the same holds for the evolution of the economy, legal systems, social systems, and culture. Because I have discussed this before with my colleagues Longo and Montevil (1,2) and elsewhere, (3,4), my discussion of this first point will be rather brief. 2) What I shall choose to call, after Aristotle's four causes, noted below, Formal Cause Laws derived from specific "ensemble theories" tell us about the world. But Formal Cause Laws are not reducible to efficient cause entailing laws of the Newtonian Paradigm and, critically, have already, unnoticed, crept into biology concerning the origin of life, and economics concerning economic growth. Formal cause laws appear to be a new way to do science, independent of efficient cause entailing laws. Thus Formal Cause laws can be independent of any specific material substrate. This may bear on the sufficiency of Materialism in our account of the world.

The intent of this article is to expand how we know our living world and what we cannot know but must live in it not knowing, and explore new ways to do science.

This article is organized as follows: In section 1 I briefly discuss Aristotle's four causes. In PART I, starting in section 2, I discuss Newton's triumph formulating the foundations of classical physics and how it led to Aristotelian efficient cause entailing laws that became reductionism and remains true even for deterministic chaos and quantum mechanics. In section 3 I briefly summarize why we can achieve no efficient cause entailing laws for the evolution of the biosphere. If true, this spells the end of the reductionist belief of a "Theory of Everything" in which all that arises in the universe is entailed by some fundamental theory. This is, in itself, if true, a deeply important negative result. It changes our view of the becoming of reality. In section 4 I show that no laws of motion entail the evolution of the economy. PART II, concerning Formal Cause Laws, starts in section 5 where I discuss my initial 1971 theory of the spontaneous emergence of collectively autocatalytic sets, CAS, (5 - 8), see also Dyson, (9), and major improvements by M. Steel and W. Hordijk, themselves and then with me, recently (10,11,12). I discuss why the theory of the spontaneous emergence of collectively autocatalytic sets is not reducible to efficient cause entailing laws of the Newtonian Paradigm. I then characterize this theory as a Formal Cause Law that gives us a new way beyond the Newtonian Paradigm to understand the world, here the expected origin of molecular reproduction, open to experimental testing. In section 6 I discuss two new Formal Law theories in economics that again are not reducible to efficient cause entailing laws but give us new insight into aspects of economic growth, (11 - 15). Both are testable and may expand current economic growth theory with potential practical implications. In the Concluding section I hope to evaluate the import of what I have written and place science in a broader framework than the Newtonian Paradigm and Reductionism.

PART I. ENTAILING EFFICIENT CAUSE LAWS AND THEIR FAILURE FOR THE EVOLVING BIOSPHERE.

Section 1. Aristotle's Four Causes

Aristotle famously considered the world as describable by four distinct causes, Material, Formal, Final and Efficient. I am not an expert with respect to Aristotle and hope that I capture at least approximately what Aristotle meant by these four cause. Consider a brick house and a statue. The Material causes of the house are the bricks and mortar. The material cause of the statue is the marble from which it is carved. The Formal cause of the house or statue is, roughly, what it IS for a house to be a house or a statue to BE a statue. The Final cause of the house is the blueprint or design of the house and the intent of the builder/owner to build the house. The Final Cause of the statue is the artistic inventions of the sculptor designing or creating in the process of carving, the statue within the marble. Final cause is Aristotle's famous "telos", which became teleology. The Efficient Cause of the house or statue are the acts of actually building the house by workpeople and tools, or for the statue, the actual work by the sculptor using his or her chisel and hammer.

Section 2. Newton and the Newtonian Paradigm: The Invention of Classical Physics And Entailing Efficient Cause Laws, Thence, Via Laplace, The Birth of Modern Reductionism.

No single mind (other than perhaps Darwin) so changed Western thought than did Newton with his invention of the mathematical framework of the differential and integral calculus and formulation of his three laws of motion and law of universal gravitation that founded Classical physics.

Consider as a simple example seven billiard balls rolling on a billiard table of fixed size and shape. Newton tells us to: i. Write down the current positions and momenta of all the balls, the initial conditions. Then we are to write down the boundary conditions given by the inner edges of the table off of which the balls bounce. Then we are to write down Newton's three laws of motion in differential equation form giving the "FORCES" between the balls and the boundaries of the table. Then we are to integrate these differential equations to obtain the future (and past) trajectories of the balls as the "solutions" of the differential equation forms of the laws of motion given the initial and boundary conditions. This ignores friction that can be handled later. But note crucially that "integration" is the deduction of the trajectories from the differential equations giving the forces between the balls. And again, critically, deduction is "entailment". This is the same as the Greek syllogism: All men are mortal, Socrates in a man, therefore Socrates is a mortal. The two premises of the syllogism logically "forces" the conclusion that Socrates is mortal. Thus the conclusion is "entailed" by the two premises. Similarly, the initial and boundary conditions, plus the laws of motion in differential equation form taken as premises, logically "force" the conclusion given by the deduced integrated solutions of the differential equations, so Newton gives us a physical world in which all that happens is both deterministic and entirely entailed. Nothing novel, in the sense of not being entailed, can arise.

Now what else had Newton done? He had formalized and matemathized Aristotle's "efficient cause" in his law of motion in differential equation form. Thus we can rightly say that Newton, and now the Newtonian Paradigm, gives us "efficient cause entailing laws" for the becoming of at least the physical world.

Less noticed is a third thing Newton had done: The boundary conditions of the billiard table, plus the positions and momenta of the balls, simultaneously "create" or "define" the very phase space of the billiard ball system, that is the boundary conditions "define" all "possible" positions and momenta of the billiard balls. Because of Newton's third law, the boundaries also play a causal role in the dynamics of the system as balls bounce off the walls.

Simon Pierre Laplace a bit more than a century later, generalized Newton to a view of an entailed becoming of the entire universe, the birth of modern Reductionism. Laplace proposed a giant computing system, the Laplacian "demon" in the sky. Were the demon to know at this instant all the positions and momenta of all the particles in the universe, then using Newton's laws, the demon could integrate Newton's laws of motion in differential equation form to achieve the deduced hence entailed becoming of the entire future and past (due to time reversibility of Newton's law) of the universe.

There are two features of this worth note. First, this is, of course, the birth of modern Reductionism, still seen in S. Weinberg's recent "Dreams of a Final Theory" (16), a hopefully simple equation that entails all that happens in the universe, and, of course, in the 10 to the 500th versions of today's string theory, (17).

Second, Laplace believed that the deterministic character of Newton's equations and thus entailed trajectories of all the particles meant that the entire future and past of the universe was "predictable". Here he was wrong as Poincare' showed.

In the 1890's the great French mathematician, Poincare', tried to solve the problem of three mutually gravitating objects under Newton's law of gravitation, for example the sun, earth and Venus. Newton, of course, solved the problem for two bodies, the sun and earth and achieved his triumph of predicted elliptical orbits, thus entailing Kepler's laws, that did so much to establish the profound creditability of Newton's classical physics. But Newton suspected there would be problems for three mutually gravitating objects.

Newton was right. Poincare' was the first to discover what we now know as "deterministic chaos". Here infinitesimal changes in initial conditions can lead to exponentially diverging trajectories. Since we cannot measure initial conditions to infinite accuracy, then even though the system is entirely deterministic, we face epistemological randomness. Nevertheless, the entire dynamics remains both deterministic and entirely entailed,(18,19).

General Relativity. The equations of General Relativity are entirely deterministic, so the efficient causal structure of spacetime is entirely entailed. One obtains in General Relativity, not an evolution "in time", for time has become another dimension of spacetime, but instead the entailed efficient causal structure of the universe.

Finally we come to the vast transformation of classical physics by quantum mechanics and the famous Schrodinger linear wave equation where the left hand side of the equation yields the propagating "waves" whose ontological status remains uncertain, set equal to a right hand side which is a classical potential, V. V constitutes the boundary conditions on the behavior of the propagating Schrodinger wave equation.

I make four points. 1). The Born Rule states that the probability that any one of the propagating Schrodinger equation's waves will be the one "measured" by a single quantum measurement is the square of the amplitude, or better, modulus, of that wave. I note that in all major versions of quantum mechanics, the specific outcome of a single quantum measurement is indeterminate. (In Bohm's version it is determinate, but the theory is completely non-local (20). 2) Given the Born Rule, the entire set of propagating amplitudes have probabilities that add up to 1.0, so the wave propagates, "unitarily" and, critically, this propagation of ALL the probabilities is entirely deterministic, hence the Schodinger equation entirely entails the evolution, not of a trajectory, but of a probability distribution. 3) Thus, with the possible exception of the outcomes of single quantum measurements which may be ontologically indeterminate, and if so, the outcome of any single quantum measurement is NOT entailed, quantum mechanics preserves the entailed becoming of the universe. I note that even if the outcome of any single quantum measurement is unentailed by quantum mechanics, the set of POSSIBLE outcomes is entirely entailed. 4. The classical potential, V, on the right hand side of the Schrodinger equation constitutes again the boundary conditions of the quantum system and again "creates" or "defines" the phase space of all the possible behaviors of the quantum system. However, unlike the classical boundary conditions of the billiard table which, with the positions and momenta of the balls, also "create" the classical phase space, but in classical physics via Newton's third law the billiard wall boundaries also play a causal role, in quantum mechanics the classical potential, V, which both constitutes the boundary conditions on the quantum system and "creates" the phase space of all the possible behaviors of the quantum system, plays no causal role at all. V is merely a boundary condition and, because it plays no causal role I want to say that V "enables" the behavior of the quantum system via the "enabling constraints" afforded by V as boundary conditions. We will find an unexpected analogy to this enablement by V in what I shall call an ever changing, and typically unprestatable, "Adjacent Possible" "created" by ever new adaptations that constitute enabling constraints in the evolution of the biosphere.

Reductionism. It remains the dream of many physicists that the entire becoming of the universe is entailed. The concept of "efficient cause", as noted, is clearly a mathematization of Aristotle's efficient cause in all of classical physics. This is less clear in the unitary propagation of the Schrodinger equation whose evolution is entirely determinisitic, but it is not clear what is "waving" in the Schrodinger wave equation and thus in what sense this deterministic wave equation remains an "efficient cause" entailing law. For the purposes of this discussion, however, I do not think this issue need detain us.

Thus the framework of all modern physics, from General Relativity to quantum mechanics to string theories as far as I know, remain an entailed becoming of our one universe or the multiverse. On the multiple world interpretation of quantum mechanics, (21), the universe splits at each quantum measurement. By splitting, measurement never happens. What are entirely entailed are each of the Possible universes that can split off as the unitary wave equation propagates.

In short in all of standard modern physics, classical and quantum, the Newtonian

Paradigm of an entailed becoming of the universe remains in place and has constituted since Newton are basic framework for science.

Section 3. No Entailing Laws, But Enablement In the Evolution Of The Biosphere. The title of this section is also, without capitals, the title of an article by Giuseppe Longo, Mael Montevil, two French mathematicians at the Ecole National Superieur, Paris, and myself, posted on Physics ArXhiv Jan 11, 2012, (1), and subsequently published, (2). In addition, I have myself expanded upon it in other publications, some now published, (3,4). I urge the reader to consult the Physics ArXhiv posted paper or the published version under the title of this section for the most detailed discussion of these topics relating the evolution of the biosphere to physics.

Here are the central claims:

1) In physics we can always prestate the phase space, say the seven billiard balls on a billiard table. By this prestatement we are enabled to write down, as Newton did, laws of motion plus boundary conditions, as in the case of the seven billiard balls on the billiard table. Here note, as above, a critical fact: the boundary condition of the edges of the table, plus the positions and momenta the seven balls, define or "create" the phase space of ALL the possible positions and momenta of the billiard balls. Further, without knowing ahead of time the boundary conditions of the billiard table's edges, we could not integrate Newton's laws of motion in differential equation form, even had we the initial conditions. Without the boundary conditions, we would have no mathematical model of the system.

2) In the evolution of the biosphere, the very phase space of what is possible for the behavior of the evolving system persistently changes, but does so in unprestatable ways, a claim defended just below. Thus we can write no equations of motion for the evolving biosphere. Further we cannot non- circularly define the "niche boundary" condtions on evolution, so could not integrate the equations of motion, even were we to have them, which we do not. Thus, we claim, the evolution of the biosphere is entailed by no efficient cause laws.

3) Without selection acting to achieve it, the evolving biosphere creates its ever new "Adjacent Possible set of opportunities that enable but do not cause ever new and typically unprestatable adjacent possible evolutionary pathways. Evolution, beyond selection's achievement, persistently creates its own future possibilities of becoming. This claim says that the becoming of the biosphere is not merely a web of cause and effect" but also of enabled adjacent possible "opportunities", "seized" by unentailed and typically unprestatable evolutionary innovations which constitute the "arrival of the fitter", an issue never solved by Darwin.

4) In a new analogy to the boundary conditions of classical and quantum mechanics, the ever changing actual "context" of the biosphere constitutes "enabling constraints" that as enabling constraints "create" the Adjacent space of possibilities into which evolution can become. Then that becoming creates a new specific evolutionary situation or context of actual adaptations that again act as enabling constraints creating ever new, typically unprestatable Adjacent Possible directions for evolution. This, if true, is striking. In classical and quantum physics there is a sharp distinction between prestated boundary conditions which "create" the phase space of the physical system, and the dynamical laws of the system, where the laws, the boundary conditions and the initial conditions are required to deduce the entailed behavior of the physical system. The evolution of the biosphere is unentailed and often unprestatable, and also the boundary conditions "co-mingle" as enabling constraints yielding the ever new Adjacent possibilities, or "opportunities" they "enable", that then gives rise to further evolutionary innovations, the typically unprestatable "arrival of the fitter", that then constitutes the next changed context of the becoming biosphere, which, in turn creates a new Adjacent Possible.

Here briefly are the central arguments:

i. Can you name all the uses of a screwdriver? Well, screw in a screw, open a can of paint, stab an assailant, wedge open a door, tie to a stick to spear a fish, rent spear to locals for 5% of the catch....Two features here are critical. The number of uses of a screwdriver is INDEFINITE. Further, while the integers are naturally orderable, there is no natural ordering of the uses, as above, of a screwdriver alone or with other objects or processes. But these two premises imply that there is no effective procedure or algorithm to list "all" the uses of a screwdriver, or, in general, no effective procedure to search for a new use of a screwdriver. We believe this conclusion is correct. This is the famous Frame Problem of Computer Science, (22), never solved since Turing invented his machine. We believe our brief argument above shows that there is no algorithmic solution to the frame problem.

ii. The above depends upon "use" as a primitive concept. "Use" is not a term in physics. Is it justifiable, or equally so, is the term "function" in the biological sense in which the function of the heart is to pump blood not make heart sounds, justified. In (3,4) I argue that "use" and "function" are fully justified for two reasons we need. a. Above the level of atoms, say for proteins length 200 amino acids, the universe would require 10 to the 39th power its lifetime to make all these proteins just once, even were all 10 to the 80th particles making such proteins on the Planck time scale. So the universe is vastly non- repeating, or "non-ergodic", above the level of atoms. Above the level of atoms, the universe is on a unique trajectory. Most complex things will never exist. b. Cells and organisms are

Kantian Wholes, (3,4), in which the parts exist for and by means of the whole and the whole exists for and by means of the parts. Kantian wholes "get to exist" via reproduction in the non-ergodic universe above the complexity of atoms. Then we can define the "function" or "use" of a part, say the heart, as that causal consequence which helps maintain the Kantian whole in exitence, thus the function of the heart is pumping blood, not jiggling water in the pericardial sac. "Functions" and "uses" are real in the universe for Kantian wholes.

iii. Now consider an evolving and reproducing lineage of bacteria in, say some new environment. Calling a "USE ACHIEVED" or "FUNCTION ACHIEVED" a "TASK ACHIEVED", the first thing to notice is that the dividing bacterium in its environment achieves a vast "Task Closure" in which all the tasks, say making membranes, building proteins, replicating DNA, making chemosmotic pumps, achieving cell division and partition of daughter DNA molecues, sensing and responding to its world, all occur. The critical issue is this: All that has to happen in evolution is that some molecular screwdriver in one of these evolving cells "finds a use" that enhances the fitness of that bacterium in its environment, there must be heritable variation for that "use", and Natural Selection acting, not at the level of the molecular screwdriver, but at the level of the whole cell in its world, will probably pull out the new variant with the new use of that molecular screwdriver. But this new use is typically not prestatable and also changes the very phase space of evolution where Task Closures of functionalities and, thus the phenotypes of Kantian Wholes are critical parts of the relevant variables in that phase space. That is, the phase space of functional closures Kantian Wholes in the evolving biosphere, which functions constitute the relevant features in evolution, hence its phase space, changes persistently in unprestatable ways. Thus, in general, we can write no laws of motion for the evolution of the biosphere.

iv. Further, the abiotic and biotic actual niche of the evolving bacterium cannot be stated non-circularly. The bacterium achieves task closure by causal pathways or quantum processes that pass via the environment. But we cannot prestate non-circularly that task closure for either the organism alone or its relevant Actual Niche alone. The total task closure involves both the organism and the "actual niche", but that task closure is only revealed after selection acts at the
level of the Kantian Whole in its world to pick the winners. Thus, we cannot prestate the niche boundary conditions on evolution.

v. Lacking foreknowledge of the changing phase space and the boundary conditions, we can write no laws of motion nor could we integrate them even if we had them. Evolution is therefore entailed by no laws at all. If true, the Newtonian Paradigm and Reductionism is broken at the watershed of evolving life. Negative results can be important, from the failure to find a contradiction in assuming that the parallel postulate of Euclid was false, to Poincare's negative result, to Godel's negative result. This negative result, if true, says we cannot even mathematize the evolution of Kantian Wholes in their worlds.

vi. The niche does not cause, but enables evolution. Heritable variation is typically due to quantum random mutations that are quantum measurement events, so acausal. The niche enables the evolution of organisms selected for fitter variants within it, but does not cause that evolution.

Section 4. No Entailing Laws of Motion for the Evolution of the Economy Can Be Written or Integrated

I next show that, like the evolving biosphere, no laws entail the evolution of the economy. Our discussion of the evolving biosphere showed that new uses of molecular screwdrivers, once selected, become actual features of the biosphere that act as enabling constraints to create ever new, partially unprestatable, Adjacent Possible opportunities that are then "seized" in evolution by the arrival of still new fitter organisms with new molecular screwdrivers or other features that find again new uses that enhance the fitness of these next organisms. In turn these new organisms are enabling constraints that create new, often unprestatable "empty adjacent possible niches", or opportunities that in turn enable, but do not cause, the next innovative step in evolution. Evolution, as noted above, is thus not only a web of efficient cause and effect, but of enabling constraints that create adjacent possible opportunities for further evolution that in turn creates new actualities that are again new enabling constraints creating yet again new adjacent possibilities for further evolution to "seize".

Just the same process occurs in the evolution of the economic web. I give two cases. The first may be only a story, the second true. In the first, engineers are trying to invent the tractor. They know they will need a huge engine block, find one, and mount it on a chassis, which promptly breaks. They mount the engine block on a succession of ever larger chassis all of which break. Then one engineer says, "You know, the engine block is so big and rigid, we can use the engine block itself as the chassis and hang everything off the engine block!" And that is indeed how tractors and in past times, formula racing cars, are and were made.

What had the engineer done? He had found a new, typically unprestatable USE of the engine block: Its rigidity could be used for the engine block to serve as a chassis. He had solved the Frame Problem in a new way that is not
algorithmically prestatable. Again, the number of uses of a screwdriver is indefinite and unorderable, so no effective procedure or algorithm can find all the uses of a screwdriver, or, in general, find a new use of a screwdriver. The new uses of screwdrivers and engine blocks are typically unprestatable.
But with the invention of the engine block AS the chassis, the very phase space of economic evolution changed in an unprestatable way, hence we can write no laws of motion for the evolving economy. Nor can we non-circularly prestate the niche uses, that is, the actual market for, the tractor which are the boundary conditions that enable that use but do not cause it. We typically cannot non-circularly prestate the market for a tractor independently of the tractor, until we have tried and succeeded, if we do, in finding uses for the tractor that some un-predefined market supports. Thus we neither have laws of motion nor do we know the boundary conditions, so cannot integrate the laws we do not have. Thus, no laws entail the evolution of the economy.

A second example makes the same point, as reported in the New York Times in 2011. A Japanese man in Tokyo lived in a very tiny apartment with 2000 books and a new baby. Desperate for space, he scanned his books into his iPad and sold the books, so had more space. Then he realized his new business opportunity: Many in crowded Tokyo lived in tiny apartments, and were similarly crowded by their own books. His new business consisted in scanning the books of these people into their own iPads, selling the books and taking a percentage of the sales price received as his income. Now, what were the enabling constraint complements of his new business that created this new, unprestatable, Adjacent Possible empty economic niche, or opportunity which opportunity enabled, but did not cause, his new business? Well, Tokyo is crowded so built many tiny apartments. Many living in these apartments owned books. Apple invented the iPad that was one way to scan books. What did the Japanese man do? He discovered or invented a "new use" for the iPad in the context of the above conditions, namely use the iPad to scan the books of others in tiny apartments, sell the books and take a share of the sales receipts as his income.

Note four things: In general, the new use the Japanese man found was unprestatable, as is the case for the tractor above. Second, the actual context of crowded apartments and the existence of the iPad created an Adjacent Possible opportunity, or possibility space, for a new business. Third, the invention of the "new use" for the iPad was an innovation that seized the new opportunity. Fourth, Apple, in inventing the iPad, surely had no intent to enable this new business in Tokyo. Thus, just as in the evolving biosphere new uses of molecular screwdrivers, once selected, just as enabling constraints that create new empty adjacent possible niches or opportunities without natural selection "struggling" to achieve those new empty niches create new evolutionary opportunities, so also, the growing economy creates new "actuals" or "contexts" where the new actual context is an enabling constraint that enables a new adjacent possible opportunity that can then be seized to create a new way to make a living. In turn that new way of making a living can become a new context that is an enabling constraint for the creation of yet another new adjacent possible opportunity.

All this is entirely missing in standard single sector models of economic growth. All this flourished in Silicon Valley, an innovation hub of the modern economy.

Tying the truths just noted to standard factors of economic growth, such as the role of adequate capital to fund new businesses, stable government, money, banks, budget constraints, demand, and an adequate theory of price formation when we cannot prestate all the future relevant goods and production capacities – a statement required by current theories of price formation - requires further work.

PART II. FORMAL CAUSE LAWS VIA ENSEMBLE THEORIES

A new form of "law", which is not efficient cause law, and which I will call "Formal Cause Law", has crept into biology and economics, if not beyond. The rest of this article examines four of these cases, two from biology concerning the origin of life, and two from economics related in new ways to understanding economic growth. We will see that Formal Cause Laws are not reducible to familiar efficient cause entailing laws, yet tell us about the real world and can be testable.

Because Formal Cause laws can be independent of any specific efficient causes, they seem to allow us to understand the world beyond Materialism, a large claim, and a potential new way to do science.

Section 5. A Formal Law Theory of the Origin of Life

In 1971, (6), I was interested in the massive problem of the origin of life, which obviously includes the origin of molecular reproduction. Most attention was and remains focused on template replication of polynucleotides such as DNA or RNA with their famous Watson-Crick nucleotide base pairing. My own intuition was and remains that Watson Crick base pairing is far too special chemically to be deeply requisite for molecular reproduction. If not such template replication, what might be an alternative approach? The clear thought was a single molecule that catalyzed its own formation from precursors autocatalytically, or more generally, a SET of molecules, each of which catalyzed the formation of one or more members of the set from some exogenous 'food molecules", so that the set as a "whole" was "collectively autocatalytic", CAS. So my first idea was CAS.

But how could such collectively autocatalytic sets possibly form spontaneously on the early earth if molecular reproduction indeed started on earth. I built from knowledge of the famous work of Erdos and Renyi, (23), on random graphs. Here one considers a set of N independent nodes and connects random pairs of nodes by a line, or "arc". Erdos and Reyni asked foundational questions and found astonishing answers. Define a "component" of such an undirected (because the arcs are lines, not arrows), graph as a connected set of nodes.

As more and more arcs are added, connecting random pairs of nodes, at first almost all nodes are not connected to other nodes and a few pairs of connected nodes appear, then some triplets of connected nodes emerge, then eventually a modest number of mid-sized clusters. But then magic occurs, adding a few more arcs will randomly connect nodes in different mid-sized clusters and suddenly most or all of the mid-sized clusters become interconnected into a "giant component" component in the graph.

For convenience, define the "size" of the largest component as the fraction of the N total nodes in that component, so this fraction ranges from 1/N to N/N =1.
More precisely, if one plots the ratio of arcs to nodes on the X axis and the size of the largest component, (the one connecting the largest number of nodes), on the Y axis, one obtains a curve reflecting the "average or generic properties" of a vast ensemble of all random Erdos Renyi graphs. This curve is remarkable as the above emergence of the "giant component" hints. At first, as the ratio of arcs to nodes increases from 0/N at the origin on the X axis, the size of the largest component says near 0/N. Then mid-sized clusters emerge and the curve slopes gradually away from the X axis and higher on the Y axis. Then the curve steepens sigmoidally at a specific ratio of arcs to nodes that is equal to 0.5. At just this point on the X axis, suddenly, averaged over the vast ensemble of random graphs, the "giant component" emerges as do cycles of all lengths. As the ratio of arcs to nodes increases further more and more single nodes and remaining small clusters are added to the giant component.

Two points are now essential. First as the total number of nodes and arcs increases towards infinity, the sigmoidal curve steepens to a step function. Thus the emergence of the giant component is a first order phase transition, like the water to ice phase transition. Equally important the Erdos Renyi results show us "why" the emergence of the giant component is "expected".

The results of Erdos and Renyi suggested how collectively autocatalytic sets might emerge spontaneously. Consider a system of chemicals undergoing reactions to form chemicals, fed by some outside source of "food molecules", say small organic molecules as part of an organic molecule chemical reaction system. But I wanted the molecules in the reaction system to be candidates to catalyze the very reactions in that reaction system hoping that a collectively autocatalytic set might emerge. The best known class of catalysts associated with life than and now are protein enzymes, with evidence that even small proteins, called peptides, can catalyze reactions, so I considered a reaction system of amino acids and small and large linear polymers of amino acids as small and large peptides. I sought conditions for the spontaneous emergence of peptide (and later RNA or RNA and/or peptide) collectively autocatalytic sets.

Any theory seeking the spontaneous emergence of collectively autocatalytic sets must consider classes of chemical reaction networks, sustained away from equilibrium by some source of "food molecules". In addition it must either know in fact, or posit some theory about which molecules in the reaction system catalyze which reactions. If these issues are known, one can examine any given food set and reaction system with catalysis included, and ask if the system contains a collectively autocatalytic set.

A first point: We do not know, in general, say for peptides and any class of reaction systems, which peptides catalyze which reactions. This question can be tested using libraries of stochastic peptides (or of RNA).

Then the first step is to consider chemical reaction systems. These, unlike Erdos Renyi graphs which consist only of nodes, are "bipartite graphs", with nodes representing molecules and small boxes, or squares, representing reactions among input and output molecules of each reaction. Chemical reactions can be one substrate one product reactions, two substrate one product reactions, one substrate two product reactions, two substrate two product reactions and more complex reactions.

In my initial model I considered the molecules, as noted, to be peptides and considered two reactions: ligation of two smaller peptides into one longer peptide, and cleavage of a longer peptide into two smaller peptides. I then made a simple representation of peptides of only two types of amino acids, labeled 1 and 0. Thus a model peptide is just a binary string (10010101) of some length. Next, I set a maximum length for such molecular "binary strings" to be length N "bits".

Based on this, the molecular reaction system consisted in all "bit strings", from length 1, ie the amino acids 1 and 0, to all 2 to the Nth power bit strings length N. The bipartite reaction graph considered for the moment only ligation reactions, shown as two bit string substrates represented as nodes, and black arrows leading to the "reaction box" where these two are ligated, then a single black arrow leading from the reaction box to the ligated product of the two substrate strings. The arrows above point only in the ligation direction, but in general, the model molecules can also cleave into the two initial substrates. Using only single arrows allows easy representation, but most real reactions are reversible and will flow thermodynamically in either direction toward chemical equilibrium, here ignored. In reality, many reactions occur spontaneously at some rate. I idealized these spontaneous reactions as so slow that the rate could be set to 0, denoted by the use of black arrows.

With the above, three immediate questions arise: i. As N increases, how many molecules are there in the reaction system? ii. As N increases, how many reactions, represented by the boxes, are there in the system? iii. What is the ratio of reactions to molecules in the system? The latter is the essential question. For the case of linear polymers and considering only ligations and cleavage reactions, as N, the length of the longest peptide, increases the ratio of reactions to polymers increases linearly in N.

Next I needed a model of which molecules catalyzed which reactions. I have used two models. The simplest is that each molecule as a fixed probability, P, to catalyse each reaction. In the model under consideration, each molecule is tested, with probability P, for whether it catalyzes each reaction and if "yes", a blue arrow is drawn from the catalyst to the reaction box representing the reaction catalysed. By assumption, catalyzed reactions occur "rapidly", denoted by changing the color of the arrows from black to red.

In any single realization of this model, once all the molecules have been assigned according to P which reactions they catalyze, some set of reaction arrows have become "red", and one can ask if the reaction system of catalyzed reactions contains a collectively autocatalytic set.

The "magic" occurs, in parallel to the emergence of the Erdos Renyi "Giant Component", as the size of the longest polymer, N, increases, for any fixed P, or P increases for fixed N. At some value of these parameters, because the ratio of reactions to molecules increases linearly with N, so many reactions are catalyzed that the "red arrow" catalyzed reaction subgraph forms a giant component of connected catalyzed reactions. At that point, with probability approaching 1.0, the reaction boxes connecting the molecules in this giant component all receive blue arrows from the very molecules in the giant component.

At just this point, a collectively autocatalytic set has emerged, (7-12). In short the emergence of collectively autocatalytic sets is an expected "phase transition", analogous to the emergence of the giant component in Erdos Renyi random graphs.

Two issues are of importance: The emergence of collectively autocatalytic sets is a phase transition in "typical" members of the ensemble of reaction systems.

Second, the theory tells us "why" the emergence of collectively autocatalytic sets are "expected". This last point will be important below.

These results are shown in a 1986 article, (7), and my first book, (9).

A second model of which molecules catalyze which reactions requires the catalyst molecule "match" the two substrates at a few of their terminal model amino acids where the two model peptides are to be ligated. Mismatches can be tunably allowed. Given a requisite match, then the candidate catalyst has again a fixed probability of catalyzing that reaction. Again, as the length of longest polymer, N, catalytic match site parameters, and probability of catalysis, P, given a match, are tuned, the same phase transition occurs and a phase transition to a collectively autocatalytic set occurs, (9).

Very recently, Michael Steel, a mathematician, and Wim Hordijk, a computer scientist, have extended this model, by relaxing my strict and unrealistic assumption that without a catalyst, no reaction occurs, ie occurs at 0 rate. Steel and Hordijk call their collectively autocatalytic systems RAFs (10,11). There are important improvements in these latter models. In my initial model the number of reactions any polymer catalyses increases exponentially in N, which is not chemically known to be plausible. In Steel and Hordijks' results, the number of reactions any polymer catalyses when a self reproducing RAF arises is only between 1 and 2, surely chemically plausible. I describe below other features Steel, Hordijk and I with them have found (12).

Before I discuss the status as Formal Cause Laws of the above theories about model ensembles of chemical reaction systems that become collectively autocatalytic spontaneously as an expected phase transition, I briefly discuss experimental work bearing on this model. These experiments demonstrate that Formal Cause Laws can be testable, although, as we will see, they are not reducible to efficient cause entailing laws.

In 1987, the first collectively autocatalytic set was created, consisting of two DNA hexamers, A and B, where A ligated two trimer nucleotide sequences to form hexamer B, and B ligated two timers to form A, (24). Between 1995 and 2000, a single autocatalytic peptide was created, (25), and at present a nine peptide collectively autocatalytic peptide set has been created (26). This result is fundamental, for it shows that molecular reproduction definitely does not require nucleotide by nucleotide template replication by Watson - Crick base pairing of DNA, RNA or similar molecules. In short, the cornerstone of the RNA World, that life must be based on RNA template replication, is demonstrably not necessary. Collectively autocatalysis, as I hoped in 1971, is far broader than the DNA or RNA double helix and its template replication. More recently, RNA collectively autocatalytic sets have been made, (27).

But the above CAS required the wisdom of a chemist to construct. Could such sets emerge spontaneously? Very recently it has been shown that a library of ribozyme molecules, each cut in half, can be put into a solution rich in magnesium ions. The half ribozymes can self assemble into hybrid ribozymes and then spontaneously form single autocatalytic molecules and 3,5,and 7 membered collectively autocatalytic sets. Importantly the longer collectively autocatalytic sets kinetically outcompete the single autocatalyst, (28) so collectively autocatalytic sets "win" this simple Darwinian race. This experiment, for the first time, shows the spontaneous formation of CAS. However, the molecules in the system are fragments of evolved RNA sequences, ribozymes.

We are not far, I hope, from showing that libraries of stochastic peptides, or stochastic RNA sequences, or both, can spontaneously form self reproducing collectively autocatalytic sets, as proposed in a broad early published patent of M. Ballivet and myself, meant also to discover the probability that stochastic peptides catalyze specific reactions and find drugs, (29-31).

If such collectively autocatalytic sets from libraries of stochastic peptides or RNA or both can form and be placed in budding lipid vesicles called liposomes (32), one can hope that we are not far from protocells. Many issues then arise, from inhibition of catalysis leading to complex dynamics, linking of exergonic and endergonic reactions as in real cells, enabling the capturing of energy and formation of chemical work cycles, and thus a minimal model of "agency", (33), and recent theory suggesting that CAS in budding liposomes, (34), will synchronize the reproduction of both the CAS and liposome forming, indeed, protocells. Other recent theoretical work shows that such systems can undergo at least limited open ended evolution, (35), and that cells doing work cycles may maximize a power efficiency at a specific displacement from chemical equilibrium, (36).

I now turn to the critical question of the status of the theory of the spontaneous emergence of collectively autocatalytic sets. Is this theory anything like the Newtonian Paradigm? The answer is NO. There are no laws of dynamical motion of chemical changing concentrations as a function of free energy, ie efficient causes, no initial or boundary conditions, hence no integration of the absent laws of motion to yield entailed trajectories in a chemical concentration space that exhibits molecular reproduction. No Newtonian Paradigm at all is here.
Instead, what we have considered is a vast ensemble of chemical reaction networks and possible assignments of which molecules catalyze which reaction and demonstrated by theorems (7,9) and improved theorems (10-12), that at some point in the parameter spaces considered, collectively autocatalytic sets emerge as an expected phase transition. But this theory ignores efficient causes, or better perhaps, the theory assumes that such efficient causes exist for the processes involved but is not based on those specific efficient causes at all other than that they exist. The theory is about the emergent organization of reproducing systems, whatever the efficient causes driving those processes may be. Indeed, below I will show below that the real economy is also a collectively autocatalytic set. The binary strings in the model, (1001010), can represent molecules or goods. The "boxes" can represent chemical reactions or economic production functions. So the CAS theory does not care about any specific efficient causes, it cares only about the emergent organization of self reproducing systems of objects and process transforming the objects and aiding those transformations, whatever the objects, transformations among them and aiding those transformations may be, thus however those processes may be efficiently caused.

In short, the theory of the spontaneous emergence of self reproducing organizations of objects and processes is NOT an efficient cause entailing law. Yet the theory seems to be telling us about the real world, for example, how the origin of molecular reproduction may be "expected" to have occurred as a phase transition. More the theory is testable, for example using libraries of stochastic peptides, RNA sequences or both. If this theory can be called a "law", which I think it can, it is not an efficient cause law. I propose that such laws are, borrowing Aristotle's "formal cause", Formal Cause Laws.

An essential point follows: The Formal Cause Law about the "expected" emergence of self reproducing systems of objects and processes is not reducible to any specific physical - or economic - realization. Consider a specific real chemical reaction system. Using free energies, chemical potential barriers to reactions, the lowering of such barriers by catalysis, can we use the Newtonian Paradigm of differential equations then integration to ask if such a system is a CAS? Of course we can. But NO SET of such real chemical efficient cause entailed dynamical systems constitutes the CAS theory I have discussed. This is true at least because the Formal Law CAS theory is independent of any specific efficient causes and defined for an entire ensemble of "random" chemical reaction networks let alone economies. Second, merely demonstrating that any set of real chemical reaction systems contains a collectively autocatalytic set does not yet tell us why such sets are EXPECTED. To account for the "expected" emergence of collectively autocatalytic sets we would have to invent the CAS theory itself, which is independent of any specific efficient causes in any specific chemical reaction network.

This raises another important issue: If CAS theory is right and is "Formal Cause Law", then it constitutes the "fundamental" account of the origin of molecular reproduction and thus in part the fundamental account of this aspect of the origin of life. Of course, CAS theory is not alone sufficient: It entirely leaves out the efficient cause aspects of the thermodynamics of open non-equilibrium systems, which is a necessary part of any such account, any specific knowledge about chemistry, supramolecular chemistry, dynamic kinetic stability, (37), and other topics. But if CAS theory "is" the fundamental account of why molecular reproduction can be expected to emerge, Formal Cause Laws are not reducible to Efficient cause entailing laws. They seem to be a newly seen, different kind of law. If so, then it is interesting that Schrodinger and Bohr are said, (37), to have worried that the then known laws of physics were insufficient to account for life. They hoped for new physical laws. But if the CAS Formal Law theory is right, eg as testable using libraries of stochastic peptides, then Schrodinger and Bohr were both right and wrong: They were right in that the efficient cause entailing laws of physics were insufficient explanations of why molecular reproduction would be expected to arise, but seem to be wrong in seeking the sufficient answer in yet further efficient cause entailing physical laws.

Thus, Formal Cause Laws tell us about the real world, but are not reducible to efficient cause laws. Therefore, we escape the belief from Laplace to Weinberg that the becoming of the universe is entirely explained by efficient cause entailing laws.

We have now found two ways beyond standard Reductionism, in Part 1 where no efficient cause laws entail the evolution of the biosphere and now in our first example in Part 2 of the Formal Cause Laws that explain the expected emergence of self reproducing molecular systems.

In the next section on economic models we will find that Formal Cause Laws give insight into the real world when, as in Part 1, that world's becoming is not reducible to efficient cause entailing laws.

Section 6 Two Formal Cause Laws of Economic Growth

I here discuss two aspects of economic growth. Both are missing from standard economic theories of economic growth which typically treat the economy as a single sector and then ask how capital, labor, human knowledge, growth in human knowledge, savings, investment and macroeconomic monetary policy bear on economic growth as shown typically in differential equation models.

But the economy is obviously not a single sector. It is a vast web of interlocked goods and production capacities, where goods can be complements or substitutes. It becomes an obvious question whether the very diversity of this economic web plays a role in the growth of new goods, production capacities, and economic sectors. The history of Silicon Valley, where new technologies created opportunities to form new companies that innovated with yet more new technologies is a case in point.

Subcritical and Supracritical Economies.

In the past 50,000 years, the diversity of goods and production capacities of the global economy has exploded from perhaps 10,000 to perhaps 10,000,000,000. This explosion of new goods and production capacities exemplifies supracritical economic growth of new ways to make a living. The current global, U.S, and First World economies are, taken together, supracritical, persistently creating new goods and production capacities.

By contrast, some economies are subcritical, they do not generate an increasing diversity of new goods and production capacities. Alberta Canada is an example. Alberta produces timber, oil, wheat, and beef, all predominately for export, with a small IT industry hanging off oil. It is not producing an increasing diversity of goods and production capacities, so is subcritical.
The distinction between supracritical and subcritical seems to be missing in economic growth theory.

Further, Alberta has a stable government, stable money, stable banks, an educated population, fine infrastructure and is the wealthiest Province in Canada. All these are taken by the famous Washington consensus to be, with other economic factors, essential conditions for economic growth. And Alberta is successful as an export economy, but is tethered to the world market. But obviously the conditions of the Washington consensus are not sufficient to lead to supracritical economic growth, nor therefore to an economy which creates, sells and buys its products largely internally, rather than depending heavily upon export markets.

These facts suggest we need new thinking. Indeed, these facts suggest that the diversity and complexity of an economy is related to its growth for which there is early empirical data, (38).

I now describe a Formal Cause Law theory of subcritical and supracritical economic growth and a critical phase transition between them. Plot on the X axis the diversity of renewable resources in some economy, say Alberta or the U.S. On the Y axis plot the diversity of production capacities. Now make the incredibly over-simple assumption that any production capacity has some probability, P, of "acting on" any renewable resource to generate a new good. Thus assign at random, hence from an ensemble, which production capacities act on renewable resources to create a first generation of new goods, denominated merely as numbers. Then iterate this process so the same production capacities can act on the first generation of new goods to produce a second generation, and so on.
Note that this model is entirely innocent of any specification of the efficient causes by which an abstract production function "acts on" a given abstract input good to produce a new abstract output good. In short, the theory is not dependent upon the specific efficient cause mechanisms.

This model, of which an earlier version is based on a "grammar model" specifying which goods are production capacities acting to transform input to output goods rather than random assignment of goods and production capacities, (13), has been solved analytically and numerically, (14). It exhibits a subcritical region in the X Y phase space near the origin and extending near the axis up and out the Y and X axis. A roughly hyperbolic curve demarks the boundary of the subcritical region, and beyond this "critical" phase transition boundary, the economy is supracritical, generating an explosion of new goods.

Missing from this model are budget constraints, a consumer with preferences and other aspects of a more realistic economy, discussed briefly in my first two books, (9,13).

The main point to make is this: Even this simple model demonstrates a first insight into the phase transition between subcritical and supracrtical economic growth in the diversity of goods, as the generic behavior of an ensemble of model economies. While not yet examined, more realistic models with a diversity of non-random assignments of which production capacities act on which goods, budget constraints, one or more consumers having utility functions, perhaps markets or social planners, will probably all exhibit the same phase transition between subcritical and supracritical economic growth. But this model is not an efficient cause model in the Newtonian Paradigm at all. It is, I propose, a Formal Cause Law.

I next describe a further published feature of this simple model, (15). If the economy is held to the critical hyperbolic line separating sub and supracritical behavior, new goods arise and old goods go extinct in small and large avalanches. Just such avalanches were noted years ago by Joseph Schumpeter, and are called Schumpeterian "Gales of Creative Destruction" in which old sectors die and new ones emerge. In our model, when the economy is held
critical, these Schumpeterian gales or avalanches show a characteristic distribution of many small and few large avalanches. Plotting the log of avalanche size on the X axis and the log of the number of avalanches of a given (binned) size on the Y axis, one obtains a straight line sloping down to the right at a slope of -1.5. A straight line in a log log plot is a power law. This curve is precisely critical dynamical behavior, similar to self organized criticality, (15).

Further, using surrogate data on firm death avalanches, on observes in real economies just the same power law distribution of avalanches. Therefore, while the data are sparse, even this simple Formal Cause model, that totally ignores any specific efficient causes, seems to be telling us about the real economy and its evolution, (15).

Collectively Autocatalytic Sets and the Evolution of the Real Economy

I end this section discussing an early application of the theory of collectively autocatalytic sets to the real economy. In its general form, the theory of the emergence of collectively autocatalytic sets concerns, as noted, objects, transformation among objects and mediation of those transformations by the objects themselves. The general theory is independent of both what the objects are, the transformations among them, how those transformations may occur, and how the objects may mediate the transformations. Thus it is independent of any specific efficient causes.

Consider the real economy. It consists in input goods that enter production processes and are transformed to output goods, and typically goods themselves carry out or mediate the production processes. For example, two boards and a nail are input goods to a production process consisting of a hand and a hammer, leading to the output good of two boards nailed together. But the hammer itself is a good produced by the economy. Hence the real economy is a 'generalized' collectively autocatalytic set, "fed" by renewable resources and human labor.

Does our real economy contain collectively autocatalytic sets, with the caveat that "catalysis" is a loose name for production functions or markets? Yes. Consider the invention of the gasoline automobile. It unleashed a Schumpeterian gale of creative destruction, wiping out the horse, buggy, saddle, corral, and so on as ways to make substantial livings except at the margins of a new economy. With the invention and spread of the gasoline automobile, its complements emerged: an oil and gas industry, gasoline stations, paved roads, lines on paved roads, traffic signals, traffic courts, motels, fast food restaurants, suburbia and people living in suburbia who needed cars. This is a collectively autocatalytic set at the heart of our modern economy.

The improved theory of CAS, called RAFs, by Steel and Hordijk (10,11) and all three of us (12), finds very important new features of RAFs, and thus their potential application to the economy. First, given a maximum length binary string, say length N, there is a maximum diversity of "goods" in the economy hence a maximum RAF. But within this maximum RAF are irreducible RAFs, having the property that removal of any member of the irreducible set would lead to the collapse of the irreducible set. Within an irreducible RAF there can be "dependent" chains of other reaction pathways that "hang off" the irreducible RAF, but do not sustain it. These can be economic activities such as the Japanese entrepreur above. Next, intermediate sized RAFs can be formed by the union of two or more irreducible RAFs, or the union of an irreducible RAF with a mid-sized RAF, or the union of two mid-sized RAFs. This process leading ultimately to the maximal RAF is a partially ordered set, hence there are many pathways of assembly of ever more complex RAFs from the irreducible ones. Further, if some RAF exists already, transient presence of some new "molecular species" can trigger the formation of a new enlarged RAF involving the initial RAF as a part of a growing "ecosystem" that provides a niche for the emergence of the enlarged RAF. Thus we have the beginnings of a theory of the evolution of either an ecosystem or an economy growing new adjacent sectors.

These models are NOT efficient cause entailing laws as in the Newtonian Paradigm. Allowing ourselves to call these models' results a Formal Cause Law, these laws are not reducible to any efficient cause or similar Newtonian Paradigm entailing law. Thus again, as in the case of CAS theory applied to the origin of life and here to the evolution of the economy, we have a new kind of Law, Formal Cause Law, that tells us about the real world. Indeed, just what kind of law is the

CAS or RAF theory can apply to molecular reaction systems and the economy? The theory is about objects, transformations among objects, and objects abetting or inhibiting those transformations, independent of the specific objects, transformations, or how transformations are abetted. It is a theory about the organization of processes, independent of efficient causes. These are, thus, Formal Cause laws, not efficient cause laws.

If the claim just made is true, that the CAS theory above is about objects, transformations among objects, and objects abetting those transformations, regardless of the material nature of those objects and transformations and how the transformations are "abetted", we seem, strikingly, to be finding laws, that our Formal Cause Laws are beyond materialism itself! The "answer" is not to be found in the "stuffs" and efficient causes among the stuffs, but in laws about the organization of processes whatever the stuffs and efficient causes among the stuffs. If so, then centuries of belief in materialism as the only basis of science seems wrong. This is a very large claim we shall have to examine carefully. It promises a new way of doing science and learning about the real world.

Second, as noted earlier in Part 1 about the economy, the examples of the invention of the tractor and the new Japanese book selling business show, the that new ways of making products and businesses are often not prestatable at all. So we can have no laws of motion, initial and boundary conditions and entailing Newton Paradigm laws, yet, using Formal Cause laws, we can find out about the real world evolving economy even though its detailed becoming is not even prestatable. Thus Formal Cause Laws, while presently remaining algorithmic, can both hope to tell us about the non-algorithmic becoming of the real world and also surpass the requirements that the relevant variables in the ever changing phase space of the evolving system even be prestatable.

Conclusions

This article argues that we are beyond the Reductionism of the Newtonian Paradigm that has dominated our thinking about how science should be done, in two ways. First, no efficient cause entailing laws for the evolution of the biosphere can be found. Thus, that evolution is beyond Reductionism. A fortiori, so too are economic, legal, social, cultural and historical evolution, with the caveat that we need not consider mind and responsible free will for the evolution of the biosphere but these thorny issues may and probably do bear on the evolution of the economy, legal systems, social systems, culture and history.
But second, we are no longer constrained to the Newtonian Paradigm to learn about the real world and its evolution even where that evolution is not prestatable and hence not open to efficient cause entaiing laws. A barely understood new class of laws, Formal Cause Laws, typically built upon the generic behaviors of large ensembles of systems, appear able to tell us about the origin of molecular reproduction and what the emergence of molecular reproduction is expected and about aspects of the evolution of the economy even though that evolution is not prestatable. It also begins to tell us about sub and supracritical economic growth independently of the unstated efficient causes of that growth, and the potential roles of generalized collectively autocatalytic sets in growing new sectors of the economy, again without attention to any specific efficient causes. Beyond Materialism, we may be finding a new way to do science.

It is unclear how the ideas presented in this article will develop. It is all too new. But I hope a beginning has been made.

REFERENCES

1) Longo, G., Montevil, M. and Kauffman, S. (2012) No entailing laws, but enablement in the evolution of lIfe. arXiv:1201.2069 **No entailing laws**, but enablement in the evolution ...

2) Longo, G. Montevil, M. and Kauffman, S. (2012) No entailing laws, but enablement in the evolution of the biosphere. in: Proceedings of the fourteenth international conference on genetic and evolutionary computation conference companion, pp. 1379–1392. doi -> 10.1145/2330784/2330946, also:!"##$%&&'()*+,)-./&+0#*#0-1)+2,30'45667896

3) Kauffman, S. A. (2012) Evolution Beyond Newton, Darwin, and Entailing Laws. Foreword for Beyond Mechanism: Putting Life Back Into Biology, eds Brian Henning and Adam Scharfe. Lexington Books (Rowman & Littlefield. pp. 1 - 24.

4) Kauffman, S. (2013). Evolution Beyond Newton, Darwin and Entailing Law: The Evolution of Complexity in the Biosphere. In Complexity and the Arrow of Time, edited by Charles H. Lineweaver, Paul C. W. Davies, and Michael Ruse, Cambridge University Press, forthcoming 2013. See http://www.cambridge.org/aus/catalogue/catalogue.asp?isbn=9781107027251

5) Kauffman, S. A. (1971) Cellular Homeostasis, Epigenesis and Replication in Randomly Aggregated Macromolecular Systems. Journal of Cybernetics, 1, 71 - 96.

6) Kauffman, S. A. (1986) Autocatalytic Sets of Proteins. J Theor. Biol 119, 1-24.

7) Farmer, J. D. Kauffman, S. A. and Packard N.H. (1986) Autocatalytic Replication of Polymers. Physica D 2, 50-67.

8) Kauffman, S. (1993) The Origins of Order: Self Organization and Selection in Evolution, Oxford University Press. N.Y. 708 p.

9) Freeman Dyson. (1999) The Origins of Life Cambridge University Press, Cambridge UK. 103 p. 280
10) Hordijk W. and Steel, M. (2004) Detecting autocatalytic, self-sustaining sets in chemical reaction systems Journal of Theoretical Biology 227(4):451-464.
11) Hordijk, W. Hein, J. and Steel, M. (2010) Autocatalytic sets and the origin of life Entropy 12(7):1733-1742
12) Hordijk, W, Steel, M., Kauffman, S. (2012).The Structure of Autocatalytic Sets: Evolvability, Enablement, and Emergence. Acta Biotheoretica. Volume 60, Issue 4, pp 379-392
13) Kauffman, S. (1995) At Home in The Universe. Oxford University Press, N.Y. 321 p.
14) Hanel, R. Kauffman, S. and Thurner, S. (2005) Phase Transtion in Random Catalytic Networks. Phys Rev ! 72, 03617.
15) Hanel, R., Kauffman, S, and Thurner, S. (2007) Towards a Physics of Evolution: Critical Diversity Dynmaics at the Edges of Collapse and Bursts of Diversification. Phys Rev E 76, 036110.
16) Weinberg, S. (1992) Dreams of a Final Theory. Pantheon Books. 235 p.??
17) Susskind, L. (2006) The Cosmic Landscape: String Theory and the Illusion of Intelligent Design. Little, Brown, and Co. N.Y. 398 p.
18) Poincare' H. (1902) La Science et l'Hypotehthese, Plammartion, Paris.
19) Poincare' H. in Longo Giuseppe, "Interfaces of Incompleteness"
http://www.di.ens.fr/users/longo/
20) Bohm, D. and Hiley, B.J. (1993). The Undivided Universe: An ontological interpretation of quantum theory,Routledge, London 399 p.
21) DeWitt, C. and wheeler, J. A. Eds, (1968) The Everett-Wheeler Interpretation of Quantum Mechanics, Batelle Rencontres: 1967 Lectures in Mathematics and Phsics.
22) Shanahan, M. (1997). Solving the Frame Problem. MIT Press. 294 p. 23) Erdos, P and RenyiA. (1960) On the Evolution of Random Graphs. Institute of Mathematics, Hungarian Academy of Sciences, publication No.5.
24) Sievers, D. and von Kiedrowski, G. (1987) Self - Replication of Hexadeoxynucleotide Analogues: Autocatalysis vesrsis Cross-Catalysis, Natie, 369/ 221-224 : doi: 10.1038/369221a
25) Lee, D. H., Granja, J. R., Martinez, J. A., Severin, K., and Ghadiri, M. R. (1996) "A self-replicating peptide." *Nature*, *382*, 525- 528.
http://www.nature.com/nature/journal/v382/n6591/abs/382525a0.html
26) Wagner, N. and Ashkenasy, g. (2009) Symmetry and order in systems chemistry. The Journal of Chemical Physics, 130 164911 - 16419.
27) Lam,. B.J. and Joyce, G. F. (2009) Autocatalytic aptazymes enable ligand- dependent exponential amplifciation of RNA. Nature Biotechnol. 27: 288.
28) N. Vaidya, N., Manapat, M.L.,Chen, I.A.. Xulvi-Brunet, R. E. Hayden, E.J., and Lehman, N. Spontaneous network formation among cooperative RNA

replicators. /Nature/ 491:72–77, 2012.

29) French Patent Office. "Procede d'obtention d'ADN, ARN, peptides, polypeptides, ou proteines par une technique de recombinaison d'ADN." French Patent number 863683, issued to Marc Ballivet and Stuart Alan Kauffman, dated 12/24/87 and registered as 2,579,518.

30) The Department of Trade and Industry. "Methods of Obtaining DNA,RNA Peptides or Proteins by means of a DNA Recombation Technique." Engligh Patent number 2183661, issue to Mark Ballivet and Stuart Alan Kauffman and dated 6/28/89

31) United States. "Process for the Production of Stochastically-Generated peptides, Polypeptides or Proteins Having a Predetermined Property." U.S. Patent number 5,817,483, issued to Stuart Kauffman and Marc Ballivet, 10/6/98/

32) Luisi, P. L., Stano, P., Rasi, S., and Mavelli, F. (2004) A possible route to prebiotic vesicle reproduction, Artifical Life, 10: 297-308.

33) Kauffman, S. (2000) Investigations. Oxford University Press, N.Y. 287 p.

34) Fillisetti, A, Serra A, Carletti, T, Villiani, M, and Poli, I. (2010) Non Linear protocell models: synchronization and chaos. Europhys. J. B. 77, 249

35) Fernando, C.; Vasas, V.; Santos, M. Kauffman, S. and Szathmary, E. (2012) Spontaneous Formation and Evolution of Autocatalytic Sets within Compartments, Biology Direct 2012, 7:1.

36) Aho. T., Kessli, J., Kauffman, S., Yli-Harja, O., (2012) Growth Efficiency as a Cellular Objective in Eschericia coli, Physics ArXhiv posted 3/21/12 link

37) Pross, A. (2012) What is Life? How Chemistry Becomes Biology, Oxford Universty Press, Oxford. 200 p.

38) Hauss P.K., Hausmann, R., Thurner, S. (2012) Empirical Confirmation of Creative Destruction from World Trade Data. Center for International Development at Harvard University CID Working Paper No 238